\documentclass[fleqn,10pt]{wlscirep}
\usepackage[utf8]{inputenc}
\usepackage[T1]{fontenc}

\usepackage{tgheros}

\usepackage{textcomp}
\usepackage{amsmath,amssymb}
\usepackage{setspace}
\usepackage{color}
\usepackage{multirow}
\usepackage{caption}
\title{Domain-Wall-Mediated Ultralow-Barrier Sliding and Pinning in Ferroelectric Moir\'e Superlattices Revealed by Machine Learning}

\author[1]{Jia-Wen Li}
\author[2,3,4,*]{Sheng Meng}
\author[1,+]{Xinghua Shi}
\author[1,$\ddagger$]{Jin Zhang}
\author[5]{Wei-Hai Fang}

\affil[1]{Laboratory of Theoretical and Computational Nanoscience, National Center for Nanoscience and Technology, Chinese Academy of Sciences, Beijing 100190, China}
\affil[2]{Beijing National Laboratory for Condensed Matter Physics, and Institute of Physics, Chinese Academy of Sciences, Beijing 100190, China}
\affil[3]{School of Physical Sciences, University of Chinese Academy of Sciences, Beijing 100049, China}
\affil[4]{Songshan Lake Materials Laboratory, Dongguan, Guangdong 523808, China}
\affil[5]{College of Chemistry, Key Laboratory of Theoretical and Computational Photochemistry of Ministry of Education, Beijing Normal University, Beijing 100875, China}

\affil[*]{smeng@iphy.ac.cn}
\affil[+]{shixh@nanoctr.cn}
\affil[$\ddagger$]{jinzhang@nanoctr.cn}

\begin{abstract}
\normalsize
Sliding ferroelectrics built from stacked nonpolar monolayers enable out-of-plane polarization and unconventional switching via interlayer sliding, yet the microscopic sliding dynamics remain unclear.
Using machine-learning molecular dynamics, we reveal spontaneous thermally driven interlayer sliding in ferroelectric MoS$_2$ moir\'e superlattices, with relative velocities on the order of $1$ m/s at 300 K.
Instead of rigid translation of the entire bilayer, the motion appears as a global drift of the moir\'e pattern.
Such thermally driven sliding is inconsistent with the meV/atom-scale rigid-sliding barrier.
In contrast, when constrained relaxation is allowed, the sliding proceeds along an almost barrierless pathway that directly reproduces the global drift of the moir\'e pattern.
Furthermore, sulfur vacancies trigger a sliding-to-pinning transition, with $\sim$ 0.1\% S vacancies already sufficient to convert the long-range sliding into localized oscillations.   	
Notably, these phenomena are not restricted to small twist angles, but arise generically in twisting-induced multidomain structures.
These results reveal that the sliding process is governed by a domain-wall-mediated collective reconstruction pathway with an ultralow barrier, rather than rigid layer translation, deepening the understanding of microscopic dynamics in moir\'e superlattices and sliding ferroelectrics.
\end{abstract}
\begin{document}

\flushbottom
\maketitle
\thispagestyle{empty}

\newpage\section*{Introduction}

Driven by the demand for next-generation storage devices, sliding ferroelectrics with switchable spontaneous polarization have emerged as exceptionally promising candidates \cite{Bian2024,Chen2026,Zhang2022a,Xue2025,Yang2024,Liang2025,Sun2025,Wu2021a}. 
Many sliding ferroelectrics have been reported, such as 3R-MoS$_2$ \cite{Bian2024,Baek2025}, 1T$'$-WTe$_2$ \cite{Fei2018a}, 1T$^\prime$-ReS$_2$ \cite{Wan2022}, $\beta$-InSe \cite{Hu2019}, $\gamma$-InSe \cite{Sui2023}, T$\rm_d$-MoTe$_2$ \cite{Jindal2023}, h-BN \cite{Yasuda2021}, MoS$_2$/WS$_2$ \cite{Rogee2022} and other transition metal dichalcogenides \cite{Wang2022b}.	
These materials show many advantages, including fatigue-free switching, high endurance, and compatibility with flexible and scalable nanodevices \cite{Yasuda2024,Li2024b,Yang2023a}.
Their spontaneous polarization arises from the redistribution of interlayer charges driven by symmetry breaking at the van der Waals interface \cite{Wu2021a}. 
Reversible polarization switching can be realized by manipulating the in-plane sliding of these atomic layers \cite{Weston2022}.

Despite extensive research, the microscopic dynamics of interlayer sliding during polarization reversal remains unclear.
One widely known picture views polarization switching as a collective rigid-like synchronized sliding of the two layers, offering an intuitive view for tracking local stacking evolution \cite{Wu2021a,Weston2022}.
However, experiments have shown that switching is more naturally described in terms of domain-wall (DW) motion \cite{Bian2024,Yang2024,Wang2022b}.
Recent theoretical works support this view by showing that polarization reversal proceeds through the formation and propagation of DWs, which show superlubric and soliton-like dynamics \cite{Ke2025,Shi2025,Wang2025,Fan2025}.
These results suggest that DWs are not merely static boundaries between domains, but highly active structural units that govern the sliding dynamics.	

Interestingly, a related picture has emerged in moir\'e superlattices, where atomic reconstruction typically produces distinct stacking domains separated by DWs, often referred to as solitons \cite{Aditya2026,Kim2025}.
These DW networks host ultrasoft shear and phason-like collective modes \cite{Maity2020}, which were recently observed experimentally in twisted bilayer WSe$_2$ \cite{Zhang2025}.
These ultrasoft collective modes imply a weak restoring force along DW-associated sliding coordinates, suggesting a low-energy pathway for interlayer motion.
Consistently, thermally driven moir\'e-pattern distortions and twist-angle relaxation were reported in experiments \cite{Jong2022,Hocking2024,Wang2016}.	
Meanwhile, classical molecular dynamics simulations revealed thermally driven drift of the moir\'e texture \cite{Maity2023}, while other calculations suggested that twisting can lower sliding barriers \cite{Sun2025a,Liang2025a,He2024}.
These DW-associated ultrasoft modes echo the superlubric DW dynamics recently proposed in sliding ferroelectrics \cite{Ke2025,Shi2025,Wang2025,Fan2025}, and naturally draw attention to ferroelectric moir\'e superlattices as a promising platform for exploring DW-mediated interlayer motion.
Locally, ferroelectric moir\'e superlattices resemble conventional sliding ferroelectrics in that both consist of ferroelectric domains separated by domain walls \cite{Wei2025,Guan2025}.
This twist-induced multidomain texture with an extended DW network makes them an ideal platform for amplifying and probing DW-mediated sliding dynamics.


In this work, we explore the molecular dynamics of ferroelectric MoS$_2$ moir\'e superlattices with a high-accuracy machine-learning potential (MLP).
The simulations reveal spontaneous thermally driven long-range interlayer sliding, with velocities on the order of 1 m/s.
Rather than a rigid interlayer translation, the observed motion manifests as a phason-like global drift of the moir\'e texture.
Such thermally driven long-range sliding is inconsistent with the meV/atom-scale rigid-sliding barrier, which would also destroy the moir\'e pattern.
We therefore calculated the relaxed sliding barrier by allowing full atomic relaxation while constraining the interlayer shift.
This pathway reduces the barrier by nearly two orders of magnitude to a nearly barrierless level and directly reproduces the global drift of the moir\'e pattern.
Together, these results reveal that the observed sliding arises from a DW-mediated reconstruction pathway with an ultralow effective barrier.
In addition, this nearly barrierless state is highly susceptible to defects, as nearly 0.1\% S vacancies could act as local pinning centers to convert this long-range sliding into bounded oscillations, thereby accounting for the absence of freely drifting moir\'e patterns in experiments.
Furthermore, both the ultralow-barrier sliding and defect-induced pinning are not restricted to small twist angles, but persist across a broad range of twist structures.
This shows that the essential ingredient is not a small twist angle with a large moir\'e period, but the twisting-induced multidomain structure and its DW networks.	
By investigating ferroelectric moir\'e superlattices, we show that their sliding process is governed by a DW-mediated collective reconstruction pathway with an ultralow energy barrier, rather than rigid layer translation, thereby deepening the understanding of moir\'e superlattices and sliding ferroelectrics.

\section*{Results}

\subsection*{Performance of the MLP for bilayer MoS$_2$}

\begin{figure*}[hbpt]
	\centering
	\includegraphics[width=0.96\columnwidth]{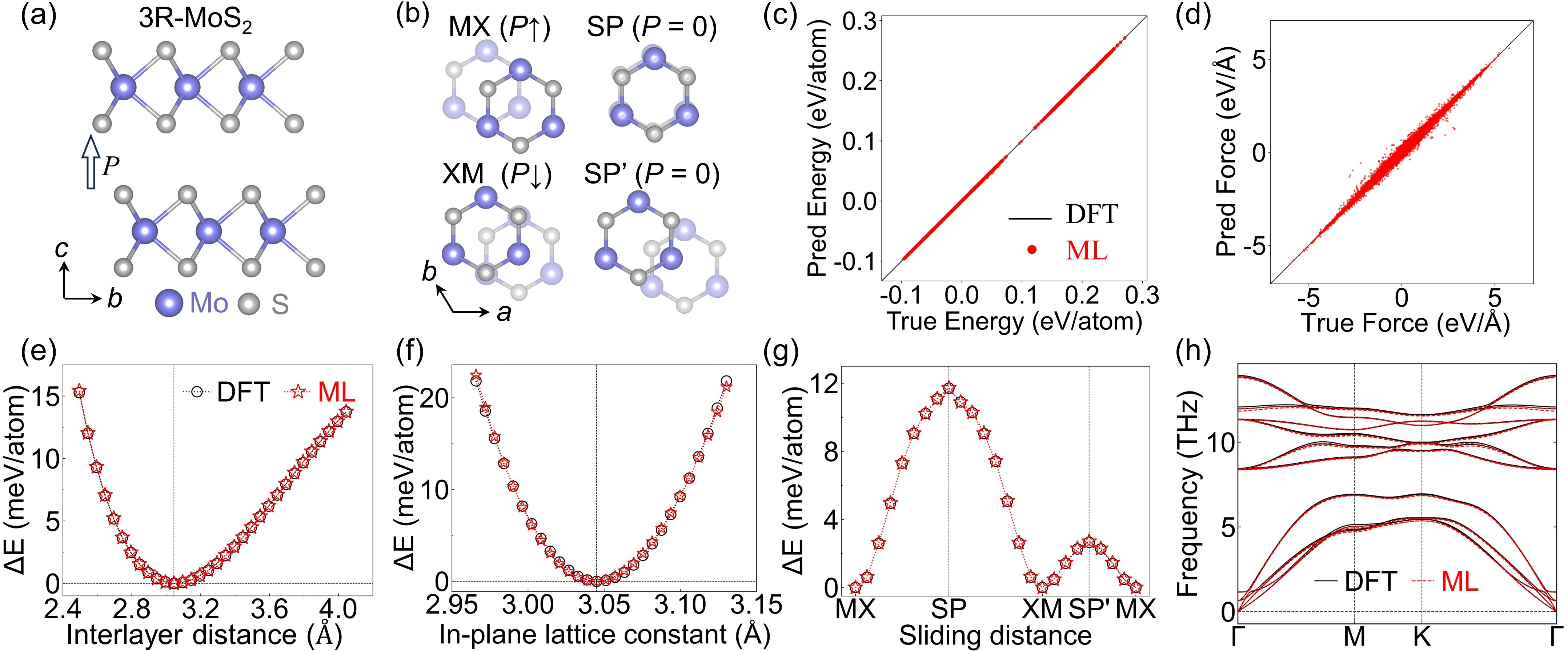}\\
	\caption{
		\textbf{Performance of the machine learning model on bilayer 3R-MoS$_2$.} 
		(a) Side view of the bilayer 3R-MoS$_2$. 
		(b) Top views of representative sliding configurations, including MX (up-polarized), SP (nonpolar), XM (down-polarized), and SP$'$ (nonpolar).
		(c, d) Comparison of the predicted and true energies (c) and forces (d), showing mean absolute errors (MAE) of 0.11 meV/atom and 13.7 meV/\AA, respectively.
		(e-g) Comparison of energies obtained by Density functional theory (DFT) and machine learning (ML) for 3R-MoS$_2$ as functions of interlayer distances (e), in-plane lattice constants (f), and sliding distances (g). 
		(h) Comparison of phonon spectra evaluated by DFT and ML.
		The results demonstrate the high accuracy of the ML model in reproducing the potential energy surface of bilayer MoS$_2$.
	}
	\label{Fig:3R}
\end{figure*}

The structure of bilayer 3R-MoS$_2$ is shown in Figure~\ref{Fig:3R}(a), whose rhombohedral stack breaks interlayer inversion symmetry and gives rise to an out-of-plane ferroelectric polarization.
Density functional theory (DFT) calculations yield an in-plane lattice constant of 3.16~\AA\ and an interlayer distance of 3.05~\AA, consistent with previous reports \cite{Bian2024}. 
Interlayer sliding modifies the stacking registry and leads to several high-symmetry configurations, including MX, XM, SP, and SP$'$, which correspond to upward polarization, downward polarization, and nonpolar states, respectively, as illustrated in Fig.~\ref{Fig:3R}(b).
The associated sliding energy profile, shown in Fig.~\ref{Fig:3R}(g), features nearly degenerate MX and XM minima separated by higher-energy nonpolar SP and SP$'$ configurations, with a sliding barrier of $\sim3$ meV/atom.

To facilitate large-scale simulations of twist-MoS$_2$ (TW-MoS$_2$), we developed a MLP based on an E(3)-equivariant graph neural network \cite{Geiger2022,Zhong2026,Gong2023,Yu2025,Yang2024b,Liang2025a,Batzner2022,Dong2025,Wang2024a}.
This is necessary because moir\'e superlattices contain thousands of atoms, well beyond the tractable size range of DFT.
A chemically resolved edge-wise gating and a trainable anisotropic distance-decay factor were introduced to capture the anisotropic and stacking-dependent interlayer coupling.
The detailed architecture is provided in Supplementary Note 1.
The model was trained on a dataset of configurations from DFT results of molecular dynamics simulations that explicitly incorporate strain, diverse stacking configurations, variable interlayer distances, and vacancies in 3$\times$3$\times$1 supercell of 3R-MoS$_2$.
The trained model yields mean absolute errors (MAE) of 0.11 meV/atom in energy and 13.7 meV/\AA\ in force, as shown in Fig.~\ref{Fig:3R}(c) and Fig.~\ref{Fig:3R}(d), respectively, reaching high accuracy comparable to recent state-of-the-art MLPs \cite{Batzner2022,Wang2024a,Dong2025,Liu2025}.
As validation, the potential energy surfaces of bilayer MoS$_2$ were faithfully captured.
The model accurately reproduces the DFT results of energy dependence on interlayer distance, in-plane lattice constant, and sliding coordinate, as well as the phonon spectrum, as shown in Fig.~\ref{Fig:3R}(e)--(h).
Benefiting from the equivariant description of local environments and symmetry, the MLP remains highly accurate in moir\'e superlattices containing hundreds of atoms including defects, as demonstrated in Supplementary Note 2.
These results validate the robustness of the MLP, providing a firm foundation for subsequent moir\'e superlattice simulations.

\subsection*{Thermally driven long-range sliding in TW-MoS$_2$}

\begin{figure}[hbpt]
	\centering
	\includegraphics[width=0.95\columnwidth]{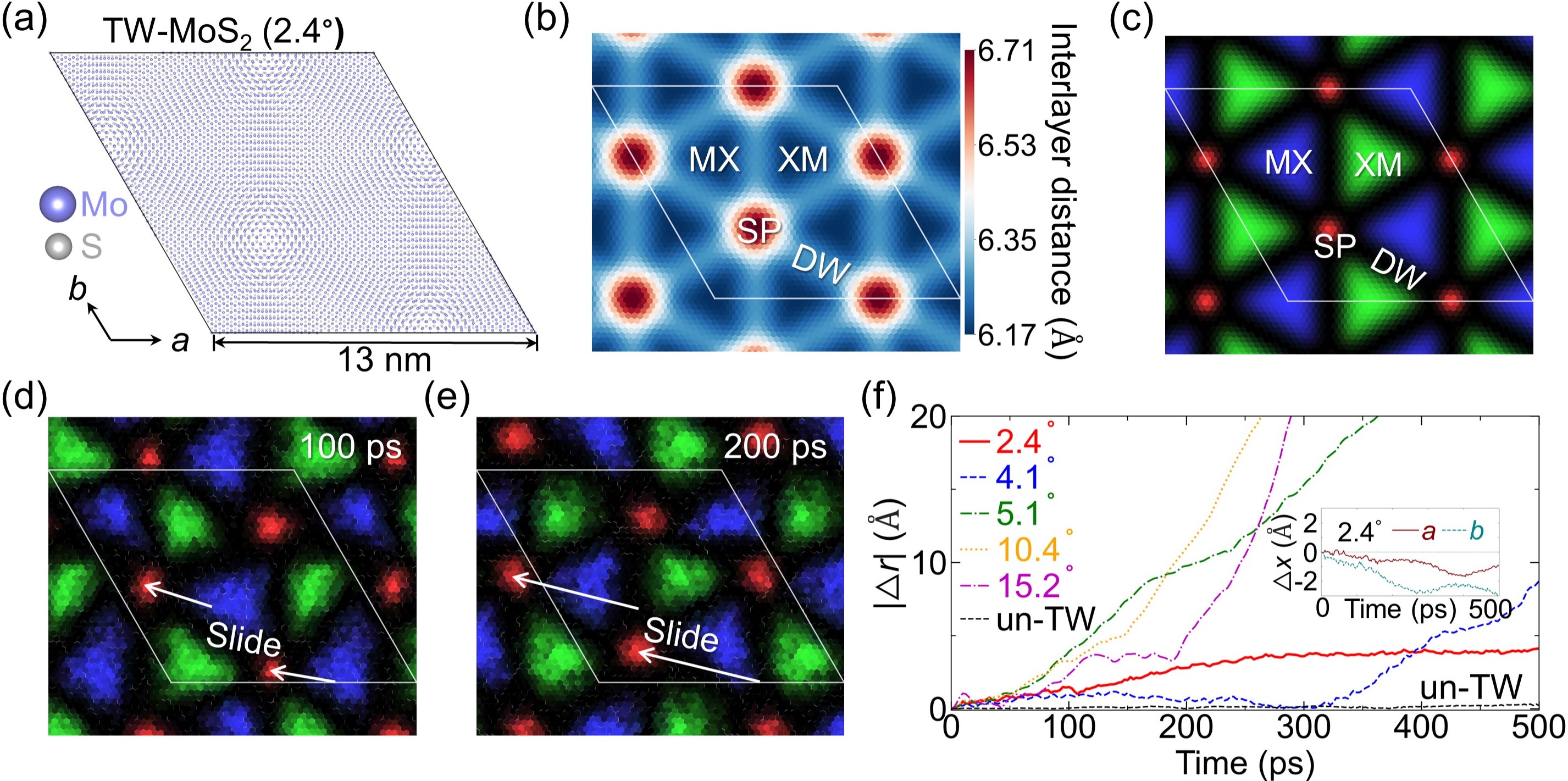}\\
	\caption{
		\textbf{Structure and thermally driven sliding of twisted 3R-MoS$_2$ with 2.4$^\circ$ twist angle.} 
		(a) Structure of the moir\'e superlattice. 
		(b) Interlayer distance map, illustrating the MX, XM, and SP domains, alongside the domain walls. 
		(c) Stack map of the moir\'e superlattice. 
		(d, e) Snapshots of stack maps at 100 ps and 200 ps obtained from machine learning molecular dynamics simulations at 300 K, showing the global motion of the moir\'e pattern. 
		(f) Interlayer relative displacement $|\Delta r|$ as a function of time, where $|\Delta r|$ denotes the magnitude of the relative center-of-mass displacement between the two layers.
		Inset shows the projected components $\Delta x$ along the $a$ and $b$ directions for the 2.4$^\circ$ structure.
		These results reveal thermally driven interlayer sliding accompanied by the global motion of the moir\'e pattern in all twisted structures, in contrast to the nearly vanishing displacement of the untwisted structure (un-TW).
	}
	\label{Fig:MD1}
\end{figure}

The relaxed TW-MoS$_2$ with a twist angle of 2.4$^\circ$ is shown in Fig. \ref{Fig:MD1}(a), featuring a large moir\'e period of 13 nm containing 10038 atoms.
Moir\'e patterns appear after structural optimization, which is performed using the trained MLP interfaced using the Atomic Simulation Environment (ASE) package \cite{HjorthLarsen2017}.
The interlayer distance map in Fig. \ref{Fig:MD1}(b) clearly displays a strongly reconstructed structure with moir\'e patterns.
Regions with smaller interlayer distances correspond to the low-energy XM/MX domains, while those with larger interlayer distances correspond to the high-energy SP regions.
This reconstructed pattern is further confirmed by the local stack map in Fig. \ref{Fig:MD1}(c), which is obtained by identifying the positions of the upper-layer Mo atoms relative to the closest Mo, S, and hollow sites at the lower layer.
Owing to the energetic preference for MX/XM over SP stacking, the SP regions are strongly compressed by the surrounding MX/XM domains during reconstruction, thereby forming a triangular network separated by DWs. 
The resulting moir\'e pattern and its triangular-domain distribution are similar to those reported previously for moir\'e superlattices \cite{Aditya2026,Guan2025,Fan2025,Kim2025,Wei2025,Liu2025,Li2024c,Maity2023,Naik2018}.
The structures and stack maps of these moir\'e superlattices are provided in Supplementary Note 3.

On this basis, we performed machine-learning molecular dynamics (MLMD) at 300 K to investigate the dynamical behavior of the moir\'e lattice with the ASE package \cite{HjorthLarsen2017}. 
As shown in Fig.~\ref{Fig:MD1}(d) and Fig.~\ref{Fig:MD1}(e), the moir\'e texture of TW-MoS$_2$ at 2.4$^\circ$ undergoes a thermally driven phason-like drift over 100 ps and 200 ps while preserving its overall topology.
Although similar thermally driven moir\'e drift has been reported in previous classical molecular dynamics simulations \cite{Maity2023}, our results further reveal a substantial global interlayer sliding underlying this moir\'e motion.
The relative interlayer center-of-mass displacement reveals substantial long-range sliding, suggesting that this interlayer sliding directly drives the moir\'e texture motion, as illustrated in Fig.~\ref{Fig:MD1}(d).
The inset further resolves the $a$- and $b$-direction components of the displacement.
The corresponding interlayer sliding velocity is on the order of 1 m/s, which is amplified into a moir\'e-texture drift velocity on the order of 40 m/s.
This motion is entirely thermally driven, because the initial interlayer relative velocity is set to zero, which also avoids the flying-ice-cube problem \cite{Harvey1998}.
No fixed directional selectivity is observed, as the sliding directions are sensitive to random seeds and temperatures, as detailed in Supplementary Note 5 .
The weak thermal driving and long-range sliding point to an exceptionally low sliding energy barrier.

More importantly, the thermally driven long-range sliding is not restricted to small-angle moir\'e superlattices with well-defined domain structures.
As shown in Fig.~\ref{Fig:MD1}(f), similar long-range drift is also observed in twisted structures with larger twist angles, such as 10.4$^\circ$ and 15.2$^\circ$, where well-defined small-angle moir\'e patterns are no longer present.
By contrast, the corresponding displacement remains nearly zero in the untwisted bilayer.
These results indicate that the observed sliding is not a special property of a specific small-angle moir\'e geometry, but arises more generally from twisting-induced multidomain structures.

\subsection*{DW-motion-mediated barrier lowering}
\begin{figure}[hbpt]
	\centering
	\includegraphics[width=0.77\columnwidth]{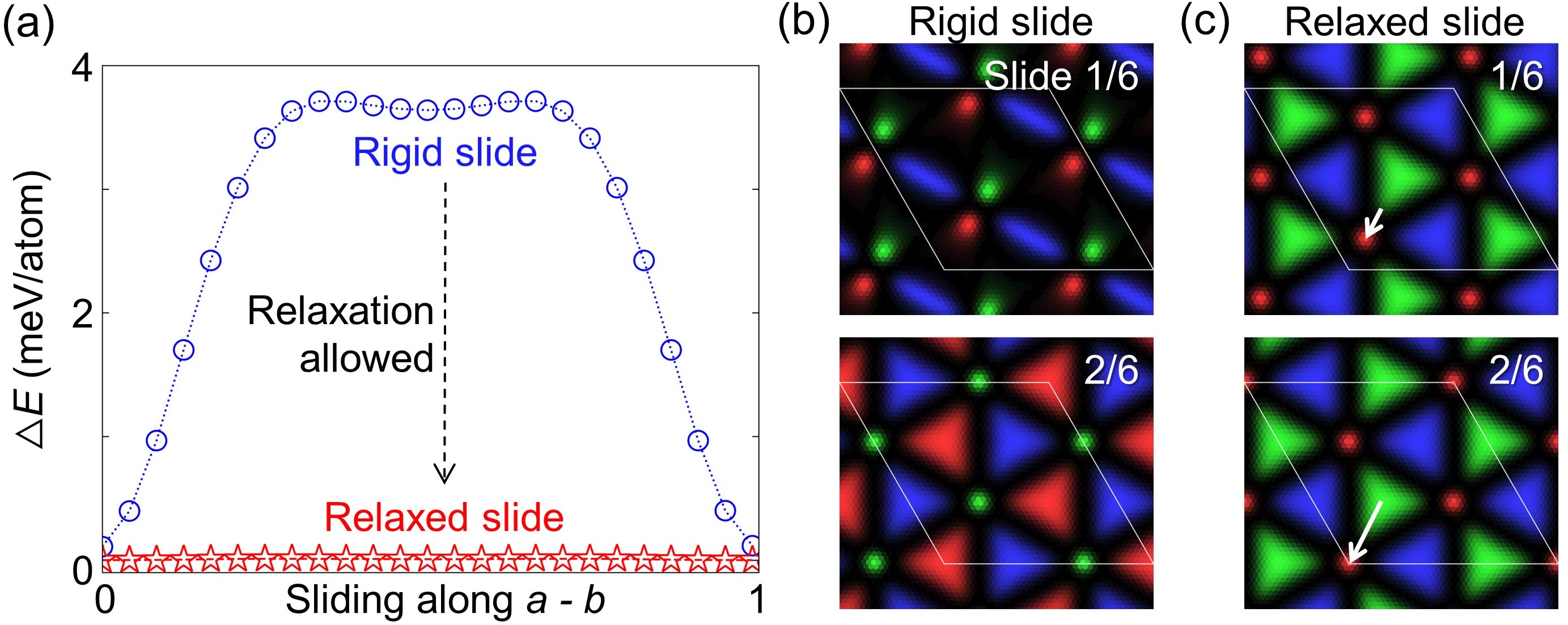}\\
	\caption{
		\textbf{Rigid versus relaxed sliding barriers in TW-MoS$_2$.}
		(a) Energy profiles along the sliding path parallel to the $a-b$ direction, comparing the rigid-sliding method with the constrained-relaxation method.
		(b) Stacking maps from rigid sliding after sliding by 1/6 and 2/6 of the path, where the moir\'e texture is strongly distorted.
		(c) Stacking maps from the relaxed pathway, where domain wall motion preserves the moir\'e texture and translates it nearly as a whole.
		These comparisons reveal a domain-wall-mediated collective reconstruction pathway in TW-MoS$_2$, which preserves the moir\'e texture and nearly eliminates the sliding barrier.
	}
	\label{Fig:Eb}
\end{figure}

To understand the origin of the observed thermally driven long-range sliding, we evaluated the sliding energy barrier.
The rigid barrier was first evaluated by artificially displacing the upper layer while fixing the in-plane atomic coordinates during relaxation.
For TW-MoS$_2$ (2.4$^\circ$), the rigid barrier shown in Fig.~\ref{Fig:Eb}(a) is about 3.8 meV/atom, comparable to that of untwisted 3R-MoS$_2$ [Fig.~\ref{Fig:3R}(g)].
This rigid barrier sharply contrasts with the observed thermally driven sliding.	
First, its magnitude is too large to permit thermally driven sliding, as untwisted 3R-MoS$_2$ does not show this behavior.
Second, rigid sliding strongly distorts the moir\'e texture [Fig.~\ref{Fig:Eb}(b)], in contrast to the global moir\'e translation observed in MLMD.
Third, it suppresses domain reconstruction, which is one of the defining features of moir\'e superlattices.
These results indicate that DW-related atomic reconstruction must be incorporated into the evaluation of the sliding barrier.

Therefore, the relaxed sliding barrier was calculated by allowing full atomic relaxation while only constraining the interlayer shift.
By imposing zero total in-plane force on each layer, the global interlayer shift is retained, while DW motion and the associated domain reconstruction are permitted.
The resulting barrier is nearly eliminated [Fig.~\ref{Fig:Eb}(a)], yielding an almost barrierless pathway.
This ultralow barrier accounts for the thermally driven sliding.
Meanwhile, the moir\'e texture remains largely preserved and undergoes an overall translation during sliding, as shown in Fig. \ref{Fig:Eb}(c).
More detailed stack maps in the sliding process are provided in Supplementary Note 4.
Such preserved global translation points to a DW-mediated collective reconstruction pathway, in which local rearrangements at the DWs collectively generate the macroscopic shift.
These results show that the ultralow-barrier sliding in TW-MoS$_2$ is governed by a DW-mediated collective pathway rather than rigid interlayer translation.
Macroscopically, it appears as global interlayer sliding and moir\'e-pattern translation, whereas microscopically it proceeds through DW migration assisted by local atomic rearrangement.
In contrast, perfect 3R-MoS$_2$ lacks DWs and thus lacks such a low-energy collective pathway.
This picture rationalizes previous theoretical reports of ultralow-damping, superlubric-like DW motion \cite{Ke2025,Shi2025,Wang2025}, the experimentally observed fast polarization switching \cite{Bian2024,Yang2024a}, thermally driven moir\'e-pattern distortions and twist-angle relaxation \cite{Jong2022,Hocking2024,Wang2016}, providing new light on sliding ferroelectric dynamics.

\subsection*{Sliding-to-pinning transition induced by sulfur vacancies}

\begin{figure}[hbpt]
	\centering
	\includegraphics[width=0.81\columnwidth]{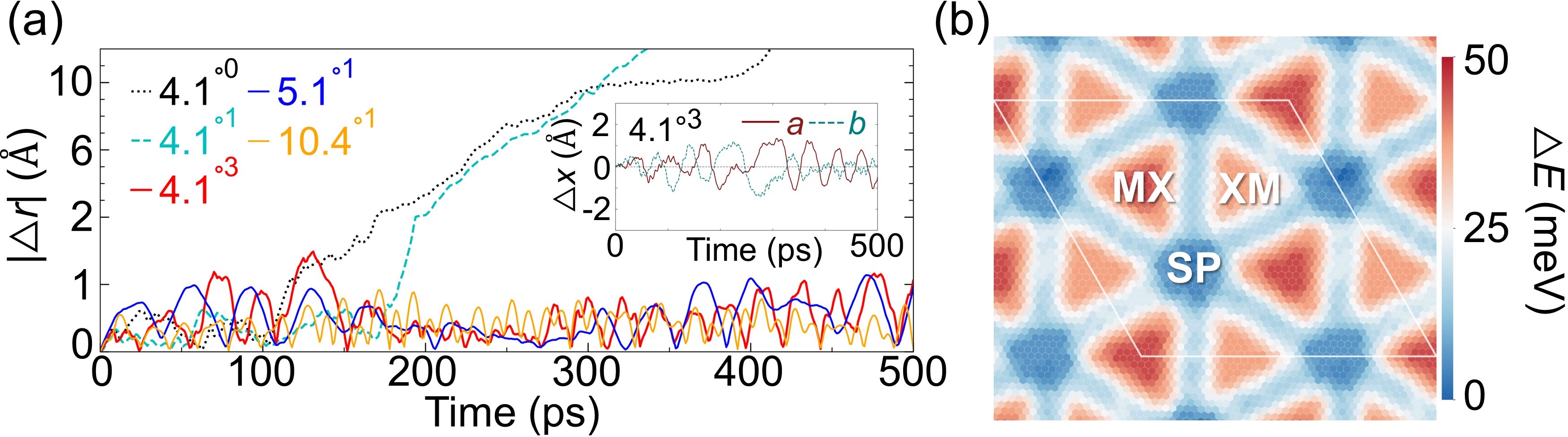}\\
	\caption{
		\textbf{Sulfur-vacancy-induced sliding-to-pinning transition.}
		(a) Time evolution of the relative interlayer displacement for the 4.1$^\circ$, 5.1$^\circ$, and 10.4$^\circ$ twisted MoS$_2$.
		Superscripts indicate the number of S vacancies.
		While the defect-free 4.1$^\circ$$^0$ and single-vacancy 4.1$^\circ$$^1$ structures still exhibit sustained sliding, bounded oscillations emerge in 5.1$^\circ$$^1$, 10.4$^\circ$$^1$, and 4.1$^\circ$$^3$, indicating pinning of the thermally driven long-range sliding.
		The inset shows the $a$- and $b$-direction components of the 4.1$^\circ$$^3$ trajectory, showing oscillation around the origin.
		(b) Relative energy landscape for a single S vacancy, showing that SP regions act as lower-energy trapping sites than the XM and MX domains.
		These results reveal that sulfur vacancies can drive a transition from long-range interlayer sliding to pinning, originating from the stacking-dependent vacancy formation energy.
		}
	\label{Fig:S}
\end{figure}

To approach more realistic experimental conditions, we next examined the effect of sulfur (S) vacancies, which are ubiquitous defects in synthesized samples \cite{Zhang2024a,Lee2021a,Zheng2026}.
Interfacial S vacancies were introduced into TW-MoS$_2$ structures with different twist angles.
The MLMD simulations show that even dilute vacancy concentrations are sufficient to suppress the thermally driven interlayer sliding and induce pinning.
As shown in Fig.~\ref{Fig:S}(a), the relative interlayer displacement exhibits a clear defect-dependent evolution.
Without defects, the 4.1$^\circ$ structure displays unbounded continuous displacement, characteristic of ultralow-barrier sliding.
By contrast, a single S vacancy already confines the 5.1$^\circ$ (0.066\%) and 10.4$^\circ$ (0.28\%) structures to bounded oscillatory motion.
Although a single vacancy does not pin the 4.1$^\circ$ structure (0.042\%), three S vacancies (0.126\%) are sufficient to arrest its long-range sliding.
The inset illustrates the pinned regime, where the interlayer trajectory oscillates around the origin rather than sliding away.
This progression confirms that sufficient defect densities generally convert long-range sliding into localized pinning under thermally driven conditions.
This picture is robust as verified in simulations with different vacancy locations and different initial random seeds, as provided in Supplementary Note 5. 
Owing to the prohibitive cost of long-time MLMD for the 2.4$^\circ$ moir\'e superlattice, the full dynamics was not explicitly resolved, whose oscillation period is expected to reach hundreds of picoseconds.

This pinning effect originates from the stacking-dependent energies of S vacancies.
Taking TW-MoS$_2$ (2.4$^\circ$) as an example [Fig.~\ref{Fig:S}(b)], a single S vacancy is energetically most favorable in the SP region, while its energy in the MX/XM regions is higher by about 50 meV.
Global moir\'e texture drift changes the local stacking configuration of the vacancy, thereby generating an additional energy barrier. 
In the dilute-defect limit, the barrier increment is expected to grow approximately with vacancy concentration.
From the crossover in the 4.1$^\circ$ case, the vacancy concentration required to pin the interlayer ultralow-barrier sliding is estimated to be about 0.1\%.
This value is already close to the experimental lower limit of sulfur-vacancy concentrations reported for MoS$_2$ \cite{Lee2021a,Zheng2026,Zhang2024a}.
Such a low threshold implies that the spontaneous thermally driven moir\'e drift is extremely vulnerable in realistic samples.
Additional extrinsic perturbations from substrates, boundaries, and other imperfections should further suppress this drift, explaining its absence in experiments.
This also explains why earlier defect-free simulations could show nearly dissipationless DW motion after field removal, unlike realistic samples \cite{Ke2025,Shi2025}.
At the same time, this pinning should not be interpreted as the complete suppression of DW motion.
Indeed, previous works have shown that electric-field-driven DW motion can remain active even in the presence of defects, which is closely connected to the fatigue-resistant and fast-switching character of sliding ferroelectrics \cite{Bian2024,He2024}.
In this sense, both the thermally driven sliding and pinned limits arise from the same DW-mediated sliding physics with ultralow barrier, further revealing that DW-related reconstruction lowers the barrier and dominates the sliding process.

\section*{Discussion}
In conclusion, our MLMD simulations reveal thermally driven long-range interlayer sliding in ferroelectric twisted 3R-MoS$_2$ moir\'e superlattices.
The sliding velocity is on the order of 1 m/s, while the moir\'e texture remains largely preserved throughout the sliding process.
This behavior is in conflict with the rigid-sliding picture, which predicts a barrier of about 4 meV/atom and would be expected to disrupt the moir\'e pattern.
Once local reconstruction is allowed under constrained interlayer displacement, however, the effective barrier is reduced by nearly two orders of magnitude, yielding a nearly barrierless pathway that directly reproduces the global drift of the moir\'e pattern.
In addition, this ultralow-barrier sliding state is strongly suppressed by interfacial S vacancies, which raise the sliding barrier because of their stacking-dependent energies.
A vacancy concentration below 0.1\% is already sufficient to drive a transition from sliding to pinning.
More generally, both ultralow-barrier sliding and pinning behaviors are not restricted to small twist angles with strong reconstruction around 2$^\circ$, but persist across a broad range of twisted structures, from the transitional regime at 4.1$^\circ$ and 5.1$^\circ$ to the near-rigid 15.2$^\circ$ case, where lattice reconstruction becomes minimal, as provided in Supplementary Note 3.
This suggests that the essential ingredient is not small twist angle itself, but the twisting-induced multidomain structures, which host DW networks to reduce the sliding barrier.
Therefore, by investigating ferroelectric moir\'e superlattices, we reveal a distinct sliding pathway in which global interlayer motion is realized not through rigid layer translation, but through DW-mediated collective reconstruction with an almost vanishing effective barrier.
This mechanism provides a deeper basis for understanding the dynamics of ferroelectric moir\'e superlattices and broadens the microscopic picture of sliding ferroelectrics.


\section*{Methods}

\subsection*{Density functional theory calculation. }

All density functional theory (DFT) calculations were performed using the Vienna ab initio simulation package (VASP) \cite{Kresse1996}.
The projector-augmented wave (PAW) method was employed to describe the electron-ion interactions \cite{Bloechl1994}.
The exchange-correlation functional was described by the generalized gradient approximation (GGA) in the Perdew-Burke-Ernzerhof (PBE) formulation \cite{Perdew1996}.
In all calculations the total energy converged to 10$^{-5}$~eV and the residual force on each atom was less than 10$^{-2}$~eV/{\AA}.
For all calculations, the plane-wave cutoff energy was set to 600~eV.
The Brillouin zone (BZ) was sampled using a 4$\times$4$\times$1 $\Gamma$-centered k-point mesh for the 3$\times$3$\times$1 supercell.
To prevent interactions between periodic images, a vacuum layer of more than 20~{\AA} was applied.	
The van der Waals interactions were considered by using the DFT-D3 method \cite{Grimme2006}.

\subsection*{Construction of moir\'e superlattices}	
Twisted bilayer 3R-MoS$_2$ moir\'e superlattices were constructed using a commensurate supercell approach \cite{Liang2025a,Mandelli2019}.
To obtain a finite moir\'e supercell compatible with periodic boundary conditions, a common superlattice vector shared by the two rotated layers must be identified.
For a given pair of coprime integers $(n,m)$, the superlattice vector of the lower layer is defined as
$\vec{T} = n \vec{a}_1 + m \vec{a}_2$, where $\vec{a}_1$ and $\vec{a}_2$ are the primitive lattice vectors of monolayer MoS$_2$.
A corresponding vector $\vec{T}'$ is then defined in the upper layer, and the twist angle $\theta$ is chosen such that $\vec{T}'$ coincides with $\vec{T}$ after rotation.
In this notation, the commensurate twist angle, moir\'e multiplicity, and total number of atoms are given by
$\cos\theta = \frac{4nm-n^2-m^2}{2(n^2+m^2-nm)}$ and $N = 6(n^2 + m^2 - nm)$.
In practice, the lower and upper layers were first expanded into matching commensurate supercells, after which the upper layer was rotated by $\theta$ relative to the lower layer and the target interlayer spacing was imposed.
The two layers were then combined into a complete bilayer moir\'e supercell, a vacuum region was added along the out-of-plane direction, and periodic boundary conditions were applied.
The resulting structures were used as the initial models for subsequent structural relaxation and molecular dynamics simulations.


\subsection*{Dataset for MLP}
The dataset for the machine-learning potential was constructed from a series of VASP molecular-dynamics trajectories on a 3$\times$3$\times$1 supercell of 3R-MoS$_2$ with 54 atoms.
All trajectories were generated at 500 K in the NVT ensemble (constant temperature and volume) with a time step of 1 fs, so as to cover representative finite-temperature configurations relevant to interlayer ferroelectric sliding in bilayer 3R-MoS$_2$.
The dataset includes molecular-dynamics trajectories with different in-plane strains (biaxial 2\% tensile strain and 2\% compressive strain), different sliding configurations along the sliding path (from 0/12 to 11/12), and different interlayer distances (2.745 \AA\ and 3.945 \AA).
The corresponding defect-containing trajectories with Mo or S vacancies at different layers were also included.
For each defect-free trajectory, 1000 frames were randomly selected, whereas 150 frames were randomly selected for each defect-containing trajectory.
Finally, 66 trajectories including 23433 frames were processed.
All configurations were randomly divided into training, validation, and test sets with a ratio of 70\%, 10\% and 20\%, yielding 16403, 2343, and 4687 structures, respectively.

\section*{Acknowledgements}
The work is supported by the starting funding from National Center for Nanoscience and Technology and the National Natural Science Foundation of China (12574255).
XingHua Shi is supported by National Key R\&D Program of China (2022YFA1203200), the National Natural Science Foundation of China 12125202.
Sheng Meng acknowledges support from the Chinese Academy of Sciences (YSBR-047) and the National Natural Science Foundation of China (12450401).

\section*{Author contributions statement}

Jia-Wen Li and Jin Zhang conceived the original ideas and supervised the work. 
Jia-Wen Li performed the first-principles calculations, code development and data analysis.
All authors participated in discussing and editing the manuscripts.

\section*{Additional information}
\textbf{Accession codes}:
The data supporting the findings of this paper are available from the corresponding authors upon reasonable request.
\textbf{Competing interests}:The authors declare no competing interests.

\end{document}